\documentstyle[sprocl]{article}

\input{psfig}

\def\Journal#1#2#3#4{{#1} {\bf #2}, #3 (#4)}

\def\be{\begin{equation}}
\def\ee{\end{equation}}
\def\bea{\begin{eqnarray}}
\def\eea{\end{eqnarray}}\def\deg{\ifmmode^\circ\else$^\circ$\fi}

\def\etal{{\rm et~al.~}}
\def\ltsim{\mathrel{\hbox{\rlap{\hbox{\lower4pt\hbox{$\sim$}}}\hbox{$<$}}}}

\def\gtsim{\mathrel{\hbox{\rlap{\hbox{\lower4pt\hbox{$\sim$}}}\hbox{$>$}}}}

\def\deg{\ifmmode^\circ\else$^\circ$\fi}

\begin{document}

\title{ THE MEASURE OF  COSMOLOGICAL PARAMETERS}

\author{W. L. FREEDMAN}

\address{Carnegie Observatories, 813 Santa Barbara St., 
Pasadena,\\ CA, 91101, USA\\E-mail: wendy@ociw.edu}


\maketitle\abstracts{  New,  large, ground  and  space telescopes  are
contributing   to  an  exciting   and  rapid   period  of   growth  in
observational cosmology. The subject is  now far from its earlier days
of being  data-starved and unconstrained,  and new data are  fueling a
healthy interplay  between observations  and experiment and  theory. I
report here on the status of measurements of a number of quantities of
interest  in cosmology:  the expansion  rate or  Hubble  constant, the
total  mass-energy  density,  the  matter  density,  the  cosmological
constant  or dark  energy component, and  the total  optical background
light.  }


\section{General Background Cosmology and Assumptions}

Based on the assumption that the universe is homogeneous and isotropic
on  large  scales,  the   framework  for  modern  cosmology  rests  on
Einstein's  general   theory  of  relativity,  leading   to  the  very
successful hot  big bang or  Friedmann--Robertson--Walker cosmological
model.  The  dynamics of the  expanding universe are described  by the
Friedmann  equation, relating the  expansion rate  to the  density and
curvature of the universe:

$$ \rm H^2  = { 8\pi G \rho_m \over 3} - { k \over R^2}  +  { \Lambda \over 3}$$

\noindent
where R(t)  is the scale factor,  H=${\dot{R} \over R}$  is the Hubble
parameter (and H$_0$  is the Hubble `constant', the  expansion rate at
the present  epoch), $\rho_m$  is the average  mass density, k  is the
curvature  term, and $\Lambda$  is the  cosmological constant,  a term
which  represents  the energy  density  of  the  vacuum.  However,  as
described in  Section 3, the last  term of the  Friedmann equation may
arise  from   a  different  form  of  dark   energy  parameterized  by
$\Omega_X$,  rather  than  a  cosmological constant.   Generally,  the
matter density is expressed as  $\rm \Omega_m = 8\Pi G\rho_m / 3H_0^2$
and  the vacuum  energy density  as  $\rm \Omega_\Lambda  = \Lambda  /
3H_0^2$.   The values  of these  parameters are  not specified  by the
model, but must be determined experimentally.

Measurements of  the cosmic  microwave background (CMB)  spectrum have
provided evidence of {\bf isotropy} on large scales at the thousandths
of  a  percent  level  \cite{fixsen96}.   But,  until  very  recently,
questions about the {\bf homogeneity}  of the universe on large scales
have remained,  primarily because of  the discovery, in the  1980s, of
unexpectedly large filaments, giant walls and voids -- features having
dimensions up to the sizes of the survey volumes themselves ($\sim$200
Mpc)  \cite{gellhuch89}.  However,  in the  most recent  generation of
galaxy  surveys  (LCRS\cite{shec96},  SDSS \cite{york00},  and  2DFGRS
\cite{colless01}),  which now  extend six  to  more than  an order  of
magnitude times greater distances, out to redshifts of z$\sim$0.3, the
largest  observed  structures  no  longer rival  the  observed  survey
volumes.    What  were  for   Einstein  two   convenient  mathematical
approximations,  homogeneity and  isotropy, are  also  apparently very
good approximations to the real universe.

Preliminary results  from 2DFGRS \cite{peacock01,percival01}  and SDSS
\cite{dodel01}  are beginning  to provide  improved statistics  on the
shape of the  galaxy power spectrum. The power  spectrum, P(k), is the
Fourier transform  of the  correlation function (which  represents the
probability of finding a pair of galaxies over a random distribution);
k is the  wave number. The power spectrum can  be parameterized by the
shape  parameter,  $\Gamma \sim  \Omega_m$h,  which characterizes  the
horizon scale at the time  of matter and radiation equality.  On small
to  intermediate   scales,  the   observed  power  spectrum   is  well
represented  by   a  power  law,   but  then  turns  over   at  larger
scales. These recent, as well  as earlier, surveys are consistent with
a value  of $\Gamma  \sim$0.15--0.20.  An important  advance resulting
from these new  large surveys is that the scale  range of the observed
galaxy  power spectrum is  now beginning  to overlap  that in  the CMB
anisotropy  measurements,  which will  offer  an increasingly  precise
means of constraining a number of cosmological parameters.

\section{The Total Matter/Energy Density}

Currently, the determination of the total matter/energy density of the
universe is most accurately  achieved from measurements of the angular
power  spectrum  of  microwave  background  temperature  anisotropies.
Recent results from a number of independent groups working with either
balloon or  high--altitude mountain  experiments are finding  that the
first  acoustic peak in  the angular  power spectrum  is located  at a
multipole value  of $l  \sim$ 210.  This  result is consistent  with a
value   of   $\Omega_{total}$   very   near   unity:

$\bullet$ Boomerang\cite{nett01,deBern01}: 1.02  $^{+0.06}_{-0.03}$

$\bullet$  DASI \cite{pryke01}: 1.04  $\pm$  0.06

$\bullet$  MAXIMA  \cite{stompor01}:  0.9$^{+0.18}_{-0.16}$

\noindent
Under the  assumption that the initial density  fluctuations that gave
rise  to these  temperature fluctuations  are Gaussian  and adiabatic,
these results  provide the  most compelling evidence  to date  that we
live in  a flat  ($\Omega_{total}$ = 1.0)  universe.  The data  are no
longer restricted to the first peak alone; there is solid evidence for
additional peaks located at $l  \sim$ 540 and 840.  CBI \cite{padin01}
has provided  coverage over  the range  $l$ = 400  to 1500,  showing a
predicted sharp decrease in power  at higher multipoles.  All of these
experiments are  consistent with an adiabatic, cold  dark matter (CDM)
inflationary  model with  a  power law  initial perturbation  spectrum
index,  n$_s$,  close  to  1.   The next  generation  of  polarization
experiments will test this hypothesis further and yield information on
the gravitational wave spectrum predicted by inflation.

\section{The Cosmological Constant / Dark Energy}

For several years, something has not  been quite right in the state of
cosmology. Until  a few years  ago, the ``standard'' cold  dark matter
(sCDM)  model  with  $\Omega_m$  =  1,  h  (=  H$_0$  /  100)  =  0.5,
$\Omega_\Lambda$ = 0  was the favored model.  However,  this model has
been  found to  fall short  in a  number of  very different  ways.  In
numerical  simulations,   sCDM  fails  to   produce  the  large--scale
distribution of  galaxies, producing too little power  on large scales
\cite{efstat90}.  The  expansion ages derived for  Hubble constants of
60 or  more (most  estimates in the  recent literature) yield  ages of
less than  11 Gyr for $\Omega_m$ =  1; these ages are  lower than most
estimates for the  Galaxy \cite{chaboyer98,truran01}, which range from
12  to  15 Gyr.   And  finally,  measurements  of the  matter  density
(Section  4) have  tended to  yield values  of only  about 1/3  of the
critical density.

All of these inconsistencies are  removed by allowing a non-zero value
for the  cosmological constant,  or more precisely,  by having  a flat
universe with  a contribution from  some form of dark  energy.  Direct
evidence for a current acceleration  of the universe has come with the
observation  that  supernovae  at  high  redshifts  are  fainter  than
predicted   by  their   redshifts,  in   comparison  to   their  local
counterparts \cite{perlm99,riess98}.   Although systematic effects due
dust  or differing  chemical  compositions could  produce an  observed
difference between the high and low redshift supernovae, several tests
have  failed   to  turn  up  any  such   systematics.   The  resulting
implication is  that the universe  is now undergoing  an acceleration,
and the data  are consistent with $\Omega_m$ =  0.3, $\Omega_X$ = 0.7,
assuming  a flat universe,  where $\Omega_X$  represents some  form of
dark  energy having  a large,  negative  pressure  \cite{turner01}.


Although  the  CMB  angular  power spectrum  contains  information  on
$\Omega_{X}$, a  direct determination of $\Omega_{X}$  is not possible
from  CMB measurements alone.  This is  due to  intrinsic degeneracies
among  the cosmological parameters  which allow  models with  the same
matter  content,  but  very  different  geometries  \cite{efstbond99}.
However,  the position  of the  first peak  in the  CMB  angular power
spectrum, combined  with large--scale structure  data, can be  used to
break    these   degeneracies   and    are   also    consistent   with
$\Omega_{X}\sim$0.75 \cite{efstat01}.

In future,  it may  be possible  to measure the  expansion rate  as a
function   of  redshift  directly   \cite{loeb98}  by   measuring  the
displacement of  individual Lyman $\alpha$ lines  in quasar absorption
spectra over a time baseline of  a decade or so. The velocity shift is
predicted to be at the level of only $\sim$2 m/sec/century (see Figure
1),  beyond current  technical reach,  but of  interest for  the next
generation of 30--meter--class  ground--based telescopes being planned
for the next decade.

\begin{figure}
\psfig{figure=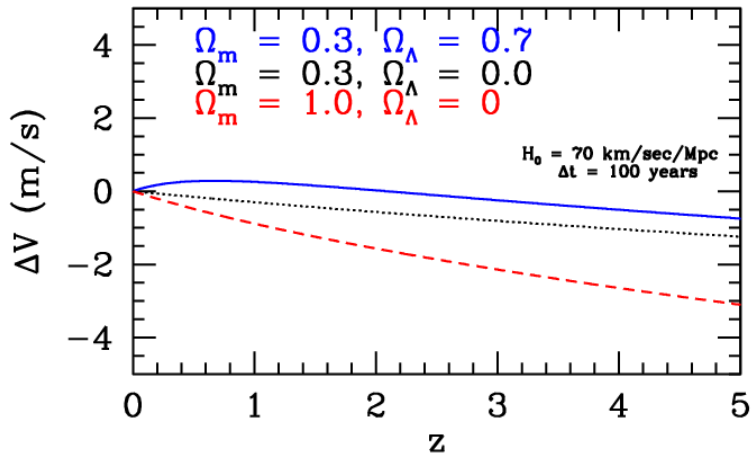,height=2.5in,angle=0}

\caption{ Predicted  velocity shift in  units of meters/second/century
as  a function  of redshift  for different  cosmologies. Based  on the
calculations  of Loeb (dashed:  $\Omega_m$ =  1, dotted:  $\Omega_m$ =
0.3, solid: $\Omega_m$  = 0.3, $\Omega_\Lambda$ = 0.7)  for a value of
H$_0$ of 70 km/sec/Mpc.  }
\end{figure}

\section{The Matter Density / Dark Matter}

There  are  several completely  independent  routes  to measuring  the
average  matter  density  of   the  universe.   In  the  past  decade,
increasingly strong evidence for dark  matter has emerged, from a wide
variety  of  independent studies.   Only  about  5\%  of the  universe
appears to  be made of ordinary  baryons.  However, there  is a strong
consensus  that the  amount of  dark matter  falls well  short  of the
critical density.

One of the most fundamental  outstanding questions in cosmology is the
nature of the dark matter.  No one simple explanation is sufficient to
explain all of the available data  - that is, there appears to be more
than  one type  of  dark matter.   There  is dark  matter in  galaxies
thought to be  in the form of ordinary baryons whose  form has not yet
been  identified, perhaps warm  (100,000\deg K)  gas or  faint stellar
remnants; there is  cold, dark matter in clusters  whose form has also
not been identified, but which  must be non--baryonic or else it would
violate  big bang  nucleosynthesis  constraints; atmospheric  neutrino
experiments yield  evidence for hot, non--baryonic dark  matter in the
form of  neutrinos with a  total mass density  about equal to  that in
luminous stars ($\ltsim$1\% of the critical density); perhaps there is
warm or self--interacting  dark matter that might explain  some of the
observed properties of galaxies on  small (galaxy) scales And there is
also the  evidence that 70\% of  the total mass/energy  density of the
universe is in a form of dark (vacuum) energy (Section 3).

Most recent estimates of the  global matter density have used clusters
of  galaxies  as probes  of  the  matter  distribution, assuming  that
clusters are large enough that  they are representative of the overall
average mass  density.  A number  of independent techniques  have been
used   for  $\Omega_m$   estimates:  cluster   mass--to--light  ratios
\cite{carlberg97},  the baryon  density  in clusters  both from  x-ray
\cite{mohr98}  and Sunyaev-Zeldovich \cite{grego01}  measurements, the
distortion  of background  galaxies  behind clusters  or weak  lensing
\cite{mellier99} (also on supercluster  scales of $\sim$3 h$^{-1}$ Mpc
\cite{wilsonkaiser01}), and the existence  of very massive clusters at
high  redshift \cite{netafan98}.  A  widespread consensus  has emerged
that  the  apparent  matter  density  appears to  fall  in  the  range
$\Omega_m \sim$ 0.2-0.4, at least on scales up to about 2h$^{-1}$ Mpc;
i.e., only $\sim$20-40\% of the  critical density required for a flat,
$\Omega_{total}$ = 1 universe.


\section{The Hubble Constant}

The Hubble  constant, the current  expansion rate of the  universe, is
one of the  most critical parameters in big  bang cosmology.  Together
with the energy  density of the Universe, it sets the  age, t, and the
size of the  observable Universe (R$_{obs}$ = ct).   The square of the
Hubble constant  relates the total  energy density of the  Universe to
its geometry.  The density of light elements (H, D, $^3$He, $^4$He and
Li) synthesized after the Big Bang also depends on the expansion rate.
Determinations   of  mass,  luminosity,   energy  density   and  other
properties  of galaxies and  quasars require  knowledge of  the Hubble
constant.   In  addition, the  Hubble  constant  defines the  critical
density of the  Universe ($\rho_{crit} = {  3 H^2 \over 8 \pi  G } $).
The critical density (and therefore H) further determines the epoch in
the Universe at which the  density of matter and radiation were equal.
Hence, the  growth of structure in  the Universe is  also dependent on
the expansion  rate.  During  the radiation era,  growth of  matter on
small  scales is  suppressed --  the  turnover in  the power  spectrum
corresponds to the point at  which the Universe changes from radiation
to matter  dominated.  This feature,  set by the critical  density, is
used to normalize cosmic structure formation models \cite{kt,pea}.

\begin{figure}
\psfig{figure=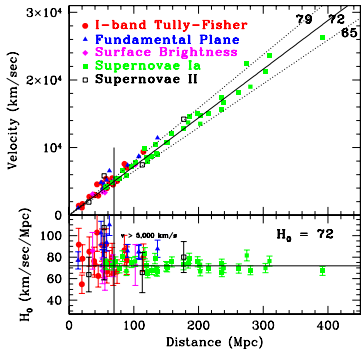,height=2.0in,angle=0}
\caption{  Velocity  versus  distance  for  galaxies  within  400  Mpc
calibrated by the Cepheid distance scale. Distances for five secondary
methods   are   plotted:  the   Tully--Fisher   relation  for   spiral
galaxies(filled  circles), type  Ia supernovae  (filled  squares), the
fundamental plane for elliptical  galaxies (filled triangles), type II
supernovae (open squares)  and surface brightness fluctuations (filled
diamonds).   A correction  for  metallicity has  been  applied to  the
Cepheid calibration. A  fit to the slope yields a value  of H$_0$ = 72
km/sec/Mpc. One-sigma error bars are also indicated.  The bottom panel
plots  H$_0$ versus distance  and the  horizontal line  corresponds to
H$_0$ = 72 km/sec/Mpc.}
\end{figure}

For several decades, over a factor of two discrepancy has persisted in
measurements of  the Hubble constant spanning  a range of  about 40 to
100 km/sec/Mpc.  Why  has the measurement of H$_0$  been so difficult?
It requires measurements of  recession velocities and the distances to
galaxies at  large distances (where deviations from  the smooth Hubble
expansion are small).  Progress in measuring H$_0$ has been limited by
the  fact  that  measuring  accurate distances  presents  an  enormous
challenge.   Primarily   as  a   result  of  new   instrumentation  at
ground-based telescopes, and with the availability of the Hubble Space
Telescope (HST),  the precision with  which H$_0$ can be  measured has
evolved at a rapid pace.


A key project of the  Hubble Space Telescope (HST) was the measurement
of distances to  a sample of nearby galaxies  using Cepheid variables.
These  intrinsically  bright  stars  follow a  well--defined  relation
between  luminosity  and period  of  variation,  from  which, given  a
measurement of apparent luminosity  and period, their distances can be
established by applying the inverse square law.  The Cepheid distances
may  then be  used to  provide  a calibration  of several  independent
methods for measuring relative distances which can be extended further
than  Cepheids: for  example,  type Ia  supernovae, the  Tully--Fisher
relation (a relation between the luminosity of a spiral galaxy and its
rotational velocity), or the fundamental plane for elliptical galaxies
(relating luminosity  to velocity dispersion). A summary  of the final
results  for the key  project \cite{wlf01,mould00}  yields a  value of
H$_0$ = 72 $\pm$ 3 (statistical) $\pm$ 7 (systematic) km/sec/Mpc based
on 5 different methods (see Figure 2).


These measurements  are consistent  with other recent  measurements of
the  Hubble constant  from the  measurement of  the Sunyaev--Zeldovich
effect  and  time delays  for  gravitational  lenses.  These  methods,
currently  with  systematics  at  the  $\pm  \sim$20--25\%  level  are
yielding values of H$_0 \sim$ 60 km/sec/Mpc \cite{reese00,schechter01}
for cosmologies in which $\Omega_m$ = 0.3, $\Omega_{\Lambda}$ = 0.7.

\section{The Optical Extragalactic Background Light (EBL)}

The luminosity of stars falls off  as ${1 \over r^2}$, but the surface
area  intercepted increases as  r$^2$.  In  an infinite  universe, the
cancelling  r$^2$  terms  ought   to  result  in  a  constant  surface
brightness, and the  sky should be as bright as the  surface of a star
(Olber's paradox).   This paradox was  resolved with the  discovery of
the  expansion of  the universe,  and  the finite  lifetimes (and  the
universe  itself).  But  the measurement  of the  actual value  of the
optical night  sky brightness has  remained elusive simply  because of
just  how {\it dark}  the sky  actually is.   And because  the optical
background light is  swamped by both by the  foreground airglow of the
Earth as well  as the foreground zodiacal light  in the ecliptic plane
(at the  level of two orders of  magnitude), to date it  has only been
possible to  place limits  on the EBL  contribution.  With HST  it has
been possible for  the first time to make a  direct measurement of the
total   optical   background    light   from   extragalactic   sources
\cite{bfm1,bfm2,bfm3}.

The total  star formation history of  the universe is  recorded in the
extragalactic background  light. The EBL  is a record of  the baryonic
mass processed  in stars, and  the formation of elements  heavier than
lithium  (the metal  production).  The  intensity of  the  EBL can  be
expressed:

$$  \bf  I_{\bf EBL}  =  {c  \over  4\pi} \int_{t_f}^{t_0}  {\rho_{\bf
bol}(t) \over 1+z} dt
 $$

\noindent
where  t$_f$ and  t$_0$  represent the  formation  and current  epochs
respectively, $\rho_{bol}$ is the total bolometric luminosity, and the
factor   (1+z)   accounts   for   the  expansion   of   the   universe
\cite{madau00,peebles93}.    The   measured   EBL  also   includes   a
contribution  from active  galactic nuclei  (AGN) and  accreting black
holes  in quasars;  recent estimates  suggest that  this contribution
could amount to about 15\% \cite{fabian99}.

An important  lower bound to  the optical luminosity  in extragalactic
sources  can   be  obtained  by  integrating   the  luminosities  from
individually  detected galaxies  (for example,  the Hubble  Deep Field
(HDF) \cite{williams96,madau00}).  However, the {\it total } amount of
light in  galaxies cannot yet  be determined directly  from individual
galaxy counts because  cosmological surface--brightness dimming (which
goes  as (1+z)$^4$)  is  sufficiently severe  that even  intrinsically
bright galaxies  of L$^*$  (Milky Way) luminosities  can be  missed at
high redshift in deep surveys, and the outer lower surface--brightness
regions  in galaxies can  easily escape  detection even  in relatively
nearby galaxies  \cite{bfm1,bfm3}.  Figure  3, from Kuchinski  {\it et
al.}  \cite{kuch01}  illustrates directly how  the well--known Messier
objects,  M51 and  M101  would appear  when  observed at  successively
higher  redshifts.  Corrections  for redshift  and surface--brightness
dimming  are  applied to  rest--frame  ultraviolet  (U--band and  1500
Angstrom)  images.   Only  the  highest  surface--brightness  features
remain visible  at high  redshift (see also  the discussion  by Colley
\etal \cite{colley97}),  and these reasonably  bright (as well  as all
fainter) galaxies will be missed in deep galaxy surveys.

\begin{figure}
\psfig{figure=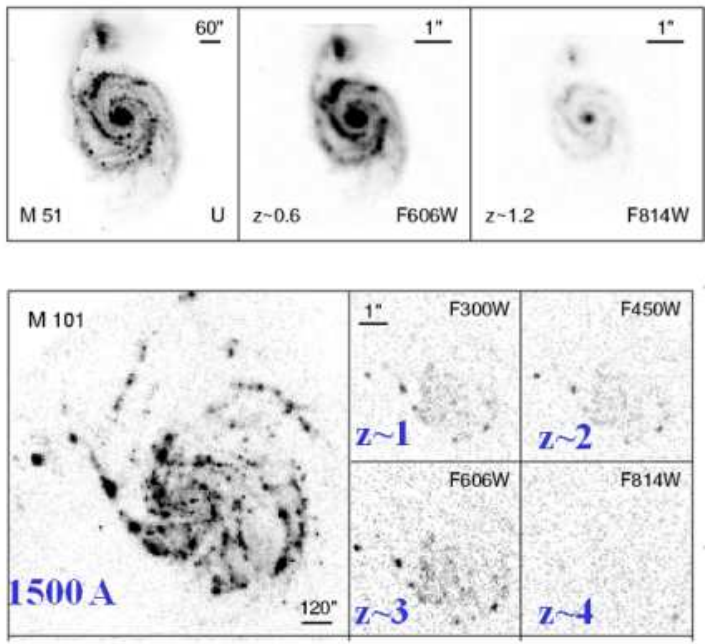,height=3.5in,angle=0}
\caption{ Ultraviolet images  of M51 and M101 as  they would appear at
successive  redshifts   from  Kuchinski  \etal.    Top  panel:  (Left)
Ground-based  U-band  (3500  Angstrom)  image of  M51.   (Center)  M51
artificially redshifted to z=0.6,  corresponding to the HST WFPC2 606W
filter.  (Right) M51 artificially  redshifted to z= 1.2, corresponding
to   the  HST   WFPC2   814W  filter.    Bottom   panel:  (Left)   UIT
far-ultraviolet  (1500  Angstrom) image  of  M101.  (Right)  Simulated
images of M101  at redshifts where the rest-frame  UIT filter bandpass
would  coincide with  the four  HST WFPC2  filters used  to  image the
Hubble Deep Field. }
\end{figure}

The  optical  HST  EBL  measurements  are  shown  in  Figure  4,  from
Bernstein, Freedman \& Madore \cite{bfm3}, in addition to measurements
at longer wavelengths.  The grey  shaded band corresponds to the model
of Dwek et  al. \cite{dwek98} scaled to fit the  HST data.  This model
is  based on  a star  formation model  including corrections  for dust
extinction and  reradiation.  From  this figure, it  can be  seen that
approximately 30\%  of the optical radiation (that  produced in stars)
is absorbed and then reradiated  by dust at infrared wavelengths.  The
total EBL contribution from 0.l to 1000 $\mu$m amounts to 100 $\pm$ 20
nW/m$^2$/sr.

\begin{figure}
\psfig{figure=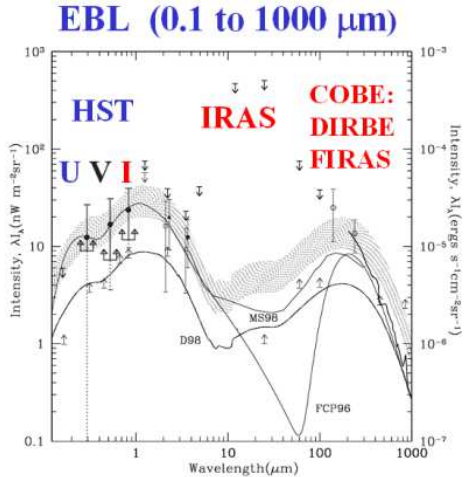,height=2.5in,angle=0}
\caption{ Background light in units of $\nu$I$_{\nu}$ from 0.1 to 1000
$\mu$m.  The  optical  (UVI)  EBL  measurements  are  from  Bernstein,
Freedman  \& Madore  (BFM) $^{\rm 37}$,  along with  EBL  measurements and
limits  at near--  and far--infrared,  and  submillimeter wavelengths;
solid  lines  are theoretical  predictions  (see  references cited  in
BFM). 
}
\end{figure}

The measured optical EBL exceeds the light computed from counts in the
Hubble Deep Field by a factor (depending on bandpass) ranging from two
to  three \cite{bfm1,bfm3}.   There are  two major  components  to the
missing  light, the  first  being  from the  outer  parts of  detected
galaxies (beyond  the aperture used for  photometry measurements), and
the second being from  galaxies below the surface brightness threshold
of the  HDF. Roughly 30\%  of the missing  light comes from  the outer
parts  of  galaxies  and   60\%  from  galaxies  below  the  detection
threshold.   This background  light measurement  is consistent  with a
star  formation  history of  the  universe  characterized  by a  steep
increase of star  formation between redshifts of 0  and 1 (as observed
in the CFRS  survey \cite{lilly96}) and a flat  or slightly increasing
star formation rate beyond in  the redshift range 1-4, consistent with
observations of  star--forming galaxies at these  redshifts (the Lyman
break galaxies)  \cite{steidel99}. Thus,  much of the  contribution to
the background light  may be due to normal  galaxies at redshifts less
than  4 that  are missed  because of  cosmological surface--brightness
dimming  (and  K-corrections  due  to band  shifting  with  redshift).
However, an earlier generation of stars, not yet detected, may also be
contributing  to the  total background  light  \cite{barkloeb00}.  The
value measured  for the  optical EBL implies  that the  mass processed
through stars contributes  about 1\% of the critical  density (for h =
0.7); i.e., a small  overall contribution, but a factor  of two higher
than previous estimates.

\section{ Concluding Remarks}

At  the current  time, there  is an  impressive convergence  on  a new
standard cosmological  model: a flat  $\Omega_{total}$ = 1  model with
$\Omega_m \sim$ 0.3, $\Omega_X \sim$ 0.7, h = 0.7, and an age of about
14  billion years.   As the  number of  independent  determinations of
these  parameters increases, so  will our  confidence in  this overall
picture.  Importantly, this convergence does not depend on the results
from a  single experiment.  For example,  a non-zero value  for a dark
energy component,  $\Omega_X$, is  implied not solely  by the  type Ia
supernovae, but  also by the combined CMB  anisotropy and large--scale
structure surveys.  The  position of the first acoustic  CMB peak at a
scale  of about  1 degree  strongly  favors a  flat universe,  whereas
several  direct  estimates  of  $\Omega_m$  yield  low  values,  again
pointing to a missing energy component.  Values of the Hubble constant
exceeding 60 km/sec/Mpc lead to a  conflict with the ages of the Milky
Way  globular clusters in  an $\Omega_X$  = 0  universe.  Many  of the
parameter measures would have to move beyond their current 2--$\sigma$
error  bounds  for  the  cosmological  model  to  change.   Given  the
difficulty in eliminating systematic errors, such a possibility cannot
be  ruled out,  but as  more accurate  data (from  MAP,  Planck, SDSS,
Chandra,  SIM,   GAIA,  LSST,  GSMT  and  other   ongoing  or  planned
experiments) become available, the prospects for robustly constraining
the  cosmological parameters, ultimately  leading to  an understanding
the physical  basis of the underlying cosmological  model, are looking
extremely promising.

\section*{Acknowledgments}
I  gratefully acknowledge  support by  NASA through  grant GO-2227-87A
from STScI  for HST observations,  and the ASTRO-2  Guest Investigator
Program through NAG8-1051 for the UIT observations.  It was a pleasure
to  participate  in  the  inauguration  of  the  Michigan  Center  for
Theoretical Physics, and  I thank the organizers for  a very enjoyable
meeting.

\end{document}